\documentclass[sigconf]{acmart}

\usepackage{amsmath}
\usepackage{balance}
\usepackage[noadjust]{cite}
\usepackage{mathptmx}
\usepackage{graphicx}
\usepackage{url}
\usepackage{array}
\usepackage{float}
\usepackage{booktabs} 
\usepackage[utf8]{inputenc}
\usepackage{multirow}
\usepackage{colortbl}
\usepackage{tikz}
\usepackage{tabto}
\usepackage{enumitem}
\usepackage[normalem]{ulem}
\useunder{\uline}{\ul}{}

\usepackage{xargs}
\usepackage{xcolor}

\usepackage[hang]{footmisc}
\usepackage{tikz}

\usepackage{pifont}

\AtBeginDocument{%
  \providecommand\BibTeX{{%
    \normalfont B\kern-0.5em{\scshape i\kern-0.25em b}\kern-0.8em\TeX}}}

\copyrightyear{2024}
\acmYear{2024}

\acmPrice{}

\usepackage{multirow}
\usepackage{pgfplots}
\pgfplotsset{compat=1.16}
\usepackage{caption} 

\begin{document}
\title[A Blockchain-based Architecture for Fake News Detection]{Establishment of a Blockchain-based Architecture for Fake News Detection}

 \author{Valdemar Vicente Graciano-Neto, Jacson Rodrigues Barbosa, Eliomar Araújo de Lima, Luiza Cintra, Rafael Medrado}
 \email{\{valdemarneto, jacson\_rodrigues,eliomar.lima\}@ufg.br}
 \email{, \{cintraluiza,rafaelcmedrado\}@discente.ufg.br}
 \affiliation{%
    \institution{Federal University of Goiás (UFG)}
    \city{Goiânia}
   \state{Goiás}
   \country{Brazil}
 }

 \author{Samuel Venzi}
 \email{samuel.venzi@goledger.com.br}
 \affiliation{%
   \institution{GoLedger}
   \city{Brasília}
   \country{Brazil}
 }

  \author{Mohamad Kassab}
 \email{mkassab@ieee.org}
 \affiliation{%
   \institution{The New York University at Abu Dhabi}
   \city{Abu Dhabi}
   \country{Arabi Emirates}
 }

\setlength\textfloatsep{\baselineskip}

 \renewcommand{\shortauthors}{Graciano-Neto, et al.}

\begin{abstract}
Fake News are a contemporary phenomenon with potential devastating effects. For inquiry and auditability purposes, it is essential that the news, once classified as false, can be persisted in an immutable means so that interested parties can query it. Although Blockchain clearly satisfies the main requirements for Fake News Management Software Systems, the prescriptive architectural solutions for that domain that cohabit Blockchain with other technologies in a single proposal still need to be made available. This paper's main contribution is presenting a prescriptive architectural solution for blockchain-based fake news management software systems. The Hoffmeister process for software architecture design is systematically followed to culminate in a software solution for that domain. The implementation of two candidate architectures and a brief simulation-based evaluation show the feasibility of the solution to satisfy the functional and quality requirements.
\end{abstract}


\maketitle


\section{Introduction}
\label{sec:introduction}

Fake News are a contemporary concern that impacts several scenarios over the world. According to the World Economic Forum, misinformation\footnote{Henceforth, terms as \textit{fake news} and \textit{misinformation} can be used interchangeably.} is at the top of global risks in 2024 \citep{WEFreport2024}. Although Artificial Intelligence (AI) models can recognize potentially false information, there is a proliferation of novel forms for creating and spreading false content. A Gartner survey with more than 200 consumers between July and August 2023\footnote{https://www.gartner.com/en/newsroom/press-releases/2023-12-14-gartner-predicts-fifty-percent-of-consumers-will-significantly-limit-their-interactions-with-social-media-by-2025} revealed that 53\% of consumers believe that the current state of social media has worsened. Furthermore, criminal acts generated from the use of \textit{DeepFake} \citep{westerlund2019emergence,yu2021survey} techniques, in which synthesized videos with audio and voice similar to those of real people, can raise the dissemination of false content to an even more worrying level. For instance while municipal elections in Brazil, that phenomenon can be even more impacting and worthy of concern.

Social medias have actually potentialized the massive dissemination of disinformation. Such dissemination can lead to catastrophic events in several areas, including personal, political, national, and corporate. Brazil is among the most significant users of digital social networks in the world\footnote{\url{https://www.forbes.com/sites/ciocentral/2013/09/12/the-future-of-social-media-forget-about-the-u-s-look-to-brazil/}}. Consequently, its users are potential targets of disinformation and, at the same time, catalysts for its dissemination. Several prior scenarios are already recognized as highly influenced by the dissemination of fake news.

In 2017, the \textit{Twitter} (currently X platform) account of Qatar's state news agency was hacked and published fake news that criticized the Arab Gulf and US Foreign Policy towards Iran. Neighboring countries such as the United Arab Emirates, Bahrain, Saudi Arabia, and Egypt broke diplomatic relations with Qatar \citep{torky2019proof}.  

Although the mainstream social media platforms, such as Instagram, X, and others, have included mechanisms for detecting and containment of fake news, doubts can be raised about the interests of the owners of those private companies. Those platforms are considered centralized and monopolized since these companies' owners integrally regulate their operations \citep{brasnamValdemar2024}. Hence, a decentralized solution, i.e., with no central authority and that could gather a diversity contributors, could overcome such problem. In that sense, blockchain emerges as a suitable solution. Decentralization is a core principle of blockchain technology because it aligns with the idea of creating systems that are open, secure, and resistant to control by any single entity \citep{namasudra2023introduction}. Based on the concepts of distributed ledger, immutability and consensus, blockchain matches the requirements of a decentralized solution and can reinforce trustability and traceability in a solution for contention of fake news dissemination in social media. Although some other solution already exist \citep{related1,related2}, they do not necessarily explore a multitude of resources brought by blockchain, including tokenization and do not involve other important resources, such as AI.

Given the importance and urgency of the topic, ANATEL (the Brazilian National Telecommunications Agency) and the Federal University of Goiás (UFG) established a Research and Development (R\&D) project to develop technologies that support the detection, classification and containment of fake news on social networks \citep{techReport}. The solutions intend to support the identification of disinformation to inform citizens what is untrue information. In that context, some requirements are essential for the solution, such as security, traceability, persistence and immutability of the information about the fake news analyzed after experts classified it. Blockchain-based approaches can satisfy all those requirements, which makes it a potential option to support the conception of the architecture for the ordered solution. 

The \textit{main contribution of this paper} is presenting a software architecture of a solution for fake news detection that, besides including blockchain infrastructure, also accommodates other cutting-edge technologies needed to accordingly deal with the complexity of the problem, as priorly illustrated. The solution advances the state-of-the-art once we are unaware of other architectural proposals that cohabitate blockchain with other technologies in a single solution. To achieve this result, we provide the methodological analysis made to achieve a minimally feasible architecture (MFA) of a solution, conceived by following the canonical principles of Hoffmeister prescriptive process for software architecture design \citep{HOFMEISTER2007}. We discuss the architecturally significant requirements, architectural alternatives, and the decisions that supported the choice of one of them. A brief architectural evaluation is also provided. 

The remainder of the paper is organized as follows: Section 2 presents a brief background on blockchain-based solutions for fake news detection and related work, Section 3 presents the research method, Sections 4, 5 and 6 show, respectively, the steps of Hofmeister's process: Analysis, Synthesis and Evaluation of the architectural design. Section 6 also discusses the results and threats to validity. Finally, Section 7 concludes the paper.

\section{Foundations on Fake News Verification and Blockchain-based Solutions}
\label{sec:background}

Fake News creates propagation bubbles (called \textit{echo chambers}) of untruths that, due to other problems such as the population's low media literacy and low critical sense to analyze the news they receive (in particular on social networks), end up feeding the imagination of its consumers, engaging them in movements that can become devastating. Furthermore, phenomena such as DeepFake \citep{yu2021survey,gomes2023}, in which videos that synthesize audio and voice similar to human people, can increase the dissemination of fake news. 

In this work, we distinguish the fake news processing steps (which we call \textit{pipeline}) into (i) monitoring, (ii) extraction, (iii) classification, and (iv) containment. These steps are essential because the main architectural elements of a technological solution for fake news address one or more of these concerns. It is worth highlighting that there may be some overlapping between the terms describing the fake news pipeline in the specialized literature, as in Shu, Bernard and Liu (2019) \citep{Shu2019}. For all intents and purposes, in this context, the main fake news processing steps are: (i) \textbf{monitoring}, which consists of the technical actions necessary to allow access to environments where fake news is often disseminated. Since many of these environments are private and owned by companies (e.g. WhatsApp, Telegram, Tiktok and Facebook), although anyone can freely access them using a free account created, access for analysis purposes is often limited. In this sense, it is necessary to obtain access, as reported by tools such as WhatsApp Monitor \citep{WhatsAppMonitor2019,reis2023}. Once you have access, the next step is \textbf{information extraction} or detection, that is, using technological resources that allow you to detect potential disseminators of fake news in addition to enabling the next steps to recognize the material produced, such as, for example, not only recognizing fake news in text, but also in audio, video and images. Once you have such information, you can proceed to the \textbf{classification} stage, using technical resources to classify news as true, false or biased. Lastly, we have the \textbf{containment} stage, in which actions can be taken to prevent or interrupt the spread of news considered false.

In the realm of blockchain-solutions for fake news containment, smart contracts play a crucial role. Smart contracts are self-executing contracts with the terms of the agreement directly written into code. They run on blockchain networks and automatically enforce and execute the terms of a contract when predefined conditions are met. They can enhance the functionality, security, and efficiency of both the consensus (voting) mechanisms, supporting the voting and applying the rules defined for the domain to compute the votes and the result \citep{mssis2024}.
\\\\
\noindent\textbf{Related Work.} Blockchain has been adopted in solutions of many domains, such as for query and registration of student degree certificates \citep{abreu2020}, healthcare \citep{Kassab21}, supply chain \citep{wei2020blockchain} and document registration service \citep{sbcars2022}. The adoption of blockchain in solutions for fake news is also not new in literature, but still a significant concern \citep{DiCicco2020,zhou2020survey,shahid:2022,kozik:2023}. 
Some existing solutions are primarily focused on the voting process to reach a consensus on the degree of fakeness of some news, working on the consensus algorithms to label the news accordingly and persist it in the blockchain. Torky et al. (2019) \citep{torky2019proof} approaches the Proof of Credibility, which establishes a formula for assessing the credibility of the source that publishes the posts; however, they value the number of followers and similar prior publications, which can be too abstract and inaccurate regarding the provenance of the source of that news. Sengupta et al. (2021) proposes the ProBlock approach  \citep{sengupta2021problock}, also based on consensus algorithms to assess the credibility of the fact-checkers, people and organizations involved in the process of classification named as \textit{fact-checking}.

Duzen et al. (2023) \citep{Duzen:2023}, for instance, present a software architecture of a tool to deal with fake news, but focusing on Social Network Analysis (SNA), not including blockchain. Kozik et al. (2023) \citep{kozik:2023} also proposes an architecture for the same domain, not including blockchain. DiCicco and Agarwal (2020) \citep{DiCicco2020} surveyed the literature to collect Blockchain Technology-Based solutions
that fight misinformation. The authors discuss nine solutions proposed by 2020, presenting their pros and cons. However, only one of them (New York Times News Provenance Project) provides architectural details, such as externalizing plug-ins and an API for developers building applications.

Other studies provide insights and highlight the relevance of \textit{blockchain-based} technologies in mitigating the challenges associated with disseminating false information. Paul et al. (2019) \citep{Paul2019} addresses the dissemination of \textit{fake news} on social networks, highlighting the importance of security in transactions and the ability of \textit{blockchain} to verify reliable sources. The use of \textit{peer-to-peer} network concepts is proposed as an effective strategy for detecting fake news in social environments. Waghmare and Patnaik (2021) \citep{Waghmare2021} proposes an innovative approach by combining machine learning and \textit{blockchain} in detecting \textit{fake news}. The creation of a \textit{blockchain} environment stands out, integrating mining, smart contracts and \textit{Proof of Work (PoW)}, with a particular emphasis on the reliability of detection through \textit{machine learning} techniques. 
Qayyum et al. (2019) propose a \textit{framework} based on \textit{blockchain} to prevent \textit{fake news}. Design issues and important considerations are discussed, offering a comprehensive look at the challenges faced in the post-truth era. Dwivedi et al. (2020) \citep{Dwivedi2020} focuses on identifying the sources of \textit{fake news} using \textit{blockchain}, presenting a framework based on \textit{blockchain} and watermarking. The ability to track the origin of fake news stands out, providing a solution to reduce its spread. Xiao, Liu and Li (2020) \citep{Xiao2020} approaches  the \textit{Internet of Vehicles} (IoV) and proposes the \textit{framework} \textit{Quick Fake News Detection} (QcFND). They integrate technologies such as \textit{Software-Defined Networking (SDN)}, \textit{edge computing} and \textit{blockchain}. 
Jing and Murugesan (2019) \citep{Jing2019} implement \textit{blockchain} on social networks to build trust in news content. The integration of \textit{blockchain} technology with advanced artificial intelligence stands out to verify the credibility of news, emphasizing the prevention of negative impacts on society.


The existing solutions are generally not accommodated in a more complex architecture (like ours). Moreover, most solutions cover only one of the steps of the fake news processing pipeline and few relevant functionalities; our solution intends to cover a more robust set of functionalities.
\section{Research Method}

Before the design of the architecture take place, a scientific workflow was firstly conducted in conformance to the following steps: (i) Exploratory literature review, for acquiring knowledge and expertise in the fake news domain, (ii) Systematic tertiary review, to systematically collect evidence from the literature (which culminated in a publication \citep{gomes2023}), (iii) Requirements elicitation meetings with the sponsor, and (iv) brainstorming. All these steps served to collect the requirements that are input for the architectural design reported in this paper, as follows. 

The research method of this study was inspired by Abreu et al. (2020) \citep{abreu2020} and involves the following steps: (i) Design of the architecture, following the systematic process proposed by Hofmeister et al. (2007) \citep{hofmeister2007general}, which involves the illustration of the typical user scenario and the assessment of candidate architectures, (ii) Development of a prototype for an application; and (iii) Evaluation, involving the \textit{Technical validation of the
prototype}, which maps the prototype characteristics with the criteria defined by Ciccio et al. (2020) \citep{di2020business}, \textit{Validation of the prototype for the applicability of the blockchain}, analyzing the viability of using blockchain for that solution through the
steps described in Pedersen et al. (2019) \citep{pedersen2019ten}; and  \textit{Prototype validation with user}, exposing a real typical user to experience the use of the prototype, and conducting interviews, with pre-established questions, aiming to compare the use of the tool with the current usage process.

\section{Architecture Design} 
\label{sec:design}

In conformance with the best practices of the state-of-the-art software architecture design (and analogously to what is conducted by Teixeira et al. (2020) \citep{TeixeiraL2020}), we followed the process proposed by Hofmeister \citep{HOFMEISTER2007} to drive the design of our software architecture. The model consists of three well-defined steps: (i) Analysis, (ii) Synthesis, and (iii) Evaluation, discussed as follows. It is important to emphasize that the requirements (functional and quality requirements)  used as input for the architectural design were collected in meetings with the Sponsor in the early moments of the project. 

\begin{figure*}[ht]
\centering
\includegraphics[scale=0.50]{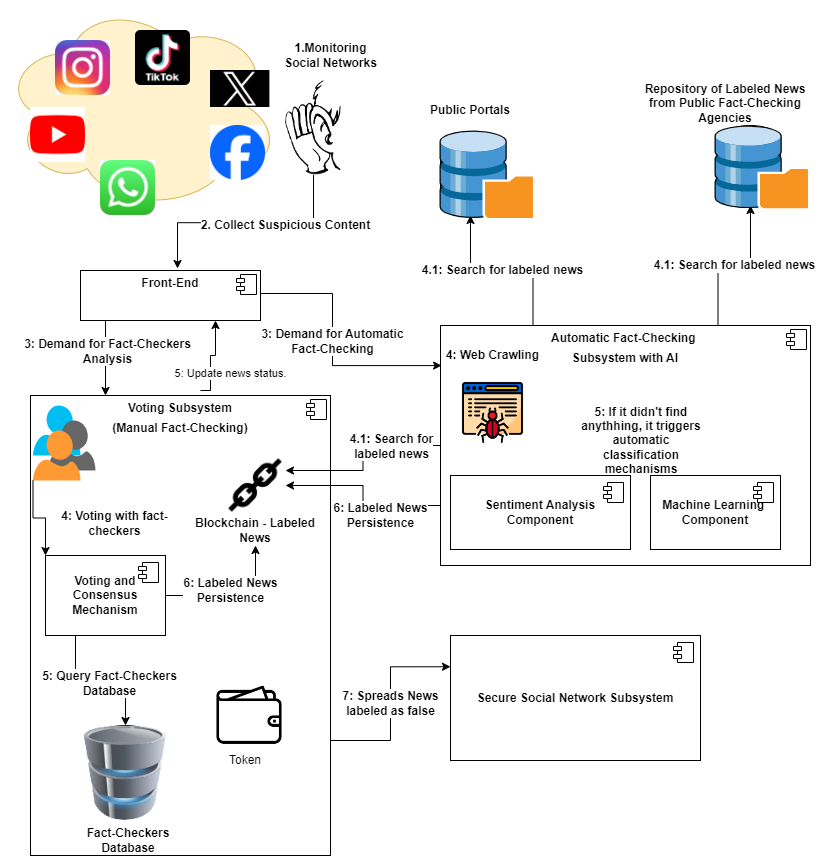}
\caption{Motivational Scenario on Fake News processing. Extracted from Private Technical Report \citep{techReport}.}
\label{fig:scenario}
\end{figure*}

\subsection{Step 1: Analysis}

This step receives, as inputs, a description of the context and the architectural concerns. As a result, the problems that the architecture solves are then found, also known as Architecturally Significant Requirements (ASR).

\subsubsection{Scenario description}

Figure \ref{fig:scenario} illustrates the motivational scenario. Ideally, a solution for fake news should have undeniable access to monitor all the news published in a diversity of social networks (e.g. X, Facebook, Instagram, TikTok), communication platforms (e.g. WhatsApp and Telegram), and portals (e.g. YouTube, News Portals and others) in a diversity of formats (text, audio, video and images) (\textbf{Step 1}). Once suspicious content is detected, it should be collected by the tool and made available for analysis (\textbf{Step 2}). The analysis (\textbf{Step 3}, called as \textit{fact-checking}) can be done under two perspectives: manual (by human fact-checkers, which can be independent freelancers or people of renowned reputation linked to fact-checking agencies) and/or semi-automatic (by AI techniques, such as sentiment analysis, information retrieval, large language models (LLM), natural language processing - NLP, and machine learning - ML). Both approaches can be performed in parallel to double-check the result (true, false or biased) before it is delivered, enhancing the confidence of the final deliberation.

\textbf{Step 4} is triggered in parallel: the semi-automatic approach endorsed by AI and the human voting supported by blockchain (the reason why there are two 'steps 4' in the figure). In the semi-automatic approach, the first step is a crawling stage (\textbf{Step 4}) to search for similar or the same news in public portals (with recognized credibility), repositories of labeled news in public agencies and also in the blockchain of the system, since that publication may have already been previously labeled. Likewise, it triggers the automatic processing mechanism (\textbf{Step 5}). This mechanism automatically extracts information from the news, recognizing typical terms and expressions in sensationalist texts. Llama LLM\footnote{\url{https://llama.meta.com/}} is also used as an oracle for that query\footnote{For deciding on Llama, an experiment was performed comparing results with ChatGPT 3.5, which will not be detailed due to space restrictions.}. Once many of those attributes are found in a single news, it receives a propensity score of falsehood, ie., a score that varies from 0 to 1; the closer the score is to 1, the more likely the news is false. Once the result is delivered, the news is labeled and persisted in the blockchain (\textbf{Step 6}), with the result also being delivered to the fact-checkers in case they are still investigating it. In \textbf{Step 7}, which is not focus of this paper, the post labeled with the result is then disseminated in the social networks towards containing/fighting the corresponding fake news dissemination.

In the manual approach, human fact-checkers analyze the suspicious news, searching for evidence that show the news is false. A voting process then takes place according to a consensus mechanism supported by smart contracts. Several consensus mechanisms exist; it can be a simple majority or a more intricate formula that expresses the credibility of the source and the effectiveness of the fact-checkers in prior judgments on fake news. Each available and involved fact-checkers vote on the likely fakeness of that news and explain/rationale that justifies/supports his/her decision. Once the consensus is reached, the decision is registered, the news is labeled and persisted in the blockchain.

In both cases (manual and semi-automatic), the result can be spread on social networks to prevent other people from believing that the news is true. 
Crypto tokens can also be implemented on top of the blockchain so that fact-checkers can be rewarded for their services in crypto assets.

\subsubsection{Architectural concerns}


Regarding the blockchain, the main objective of the system is to provide a decentralized Proof of Concept (PoC) solution that meets the following main functionalities: \textbf{offering a voting mechanism for fact checkers} and support \textbf{security, provenance and immutabiliy}. 

\subsubsection{Architecturally Significant Requirements (ASR)}

As mentioned earlier, ASR are the problems that the architecture must solve. They are a subset of the requirements that must be met before the architecture can be considered ``stable''. For each Significant Macro Requirement (SMR), respective ASRs can be derived. Next, we discuss the requirements and their respective ASRs.

\noindent\textbf{[SMR1] Interoperability} - It refers to the degree to which two or more systems, products or components can exchange information and use the information that has been exchanged \citep{ISO/IEC2010}. The consensus/voting mechanism has to be interoperable. The system should externalize an API (as reported in \citep{DiCicco2020}) for invoking services such as querying the base of news labeled as true, suspicious or false. Under this perspective, the system should be able to:
	
\begin{itemize}
            \item\textit{[ASR1] Communicate with external agencies and repositories:} The system should be capable of querying public repositories of fact-checking agencies and public portals;
            \item\textit{[ASR2] Receive request for capabilities/information}: The system should also be capable of providing an interface for queries and for other developers to create solutions, forming an ecosystem;
\end{itemize}

\noindent\textbf{[SMR2] Performance and Real time} - The response time for returning a query to the public blockchain should not exceed a given threshold. Moreover, fact-checking agencies should be able to carry out their vote as soon as there is novel suspicious news. Consensus should also take no more than a given threshold. Under this perspective, the system should cope with the following ASR:
\begin{itemize}
            \item\textit{[ASR3] Vote about the news authenticity (True, False, Partial):} The system should support the fact-checkers to vote about the likely fakeness of a publication and also include explanation/decision justification field, mentioning the source, social network analysis, and the type of manipulation, such as a content taken out of context and others. This should happen in a timely manner (couple hours, if feasible), since the impacts of the fake news dissemination increases over time. This imposes a restrict performance threshold, demanding real-time response, if possible;

\end{itemize}

\noindent\textbf{[SMR3] Scalability:} It must be possible to expand the user list to allow access to hundreds or thousands of fact checkers. Public blockchain query users can reach millions of hits and the application should be accordingly adjusted.	
\\\\
\noindent\textbf{[SMR4] Descentralized/Modularity:} In a decentralized blockchain network, no one has to know or trust anyone else. 
If a fake news vote's ledger is altered or corrupted in any way, it will be rejected by the majority of the fact-checkers in the network. The solution architecture must be modular in the sense of allowing functionality to be accordingly accommodated. \\ 
\\
\noindent\textbf{[SMR5] Integrity/Immutability and Traceability:} All news labeled and persisted on the public blockchain must be backed by those persisted on the private blockchain.
\begin{itemize}
            \item\textit{[ASR4] List Suspicious News:} The system should be capable of listing emerging publications with suspicious content;
            \item\textit{[ASR5] Dispatch news classification order}: The system should notify fact-checkers about the need to judge the veracity of one or more news and trigger the classification process in both manual and automatic approaches;
            \item\textit{[ASR6] Create a unique hash for the news:} The system should receive the news in any format (image, video, text or audio) and generate a correspondent hashing code to uniquely identify it in the system; 
            \item\textit{[ASR7] Obtain news metadata:} The system should collect relevant metadata from the news, such as creation date, content, author, source platform and others;
\end{itemize}

\noindent\textbf{[SMR6] Security:} The system should be secure for all the potential users: fact checkers, news consumers, and agencies.  
\begin{itemize}
            \item\textit{[ASR8] CRUD Fact-Checkers:} It should be possible to Create, Read, Update, and Delete fact-checkers in the system database;
            \item\textit{[ASR9] Login (for Fact Checker)}: The access to the system should only be granted under credentials;           
\end{itemize}

\noindent\textbf{[SMR7] Rewarding:} The system should encourage the participation of fact-checkers (remuneration in crypto or some other asset - currency, social asset, etc. - taking care to make the entity vote responsibly, trying to equalize the accuracy of voting with the desire to do quickly to gain more financial return with each contribution; think about the degree of reliability/credibility, etc.)

The Step 2 (Synthesis) will be shown in a separate section, since it comprises the conception of the tool itself, as follows. And Step 3 (Evaluation) will be shown in the following section.

\begin{figure*}[hbt]
  \centering
  \includegraphics[scale=0.30]{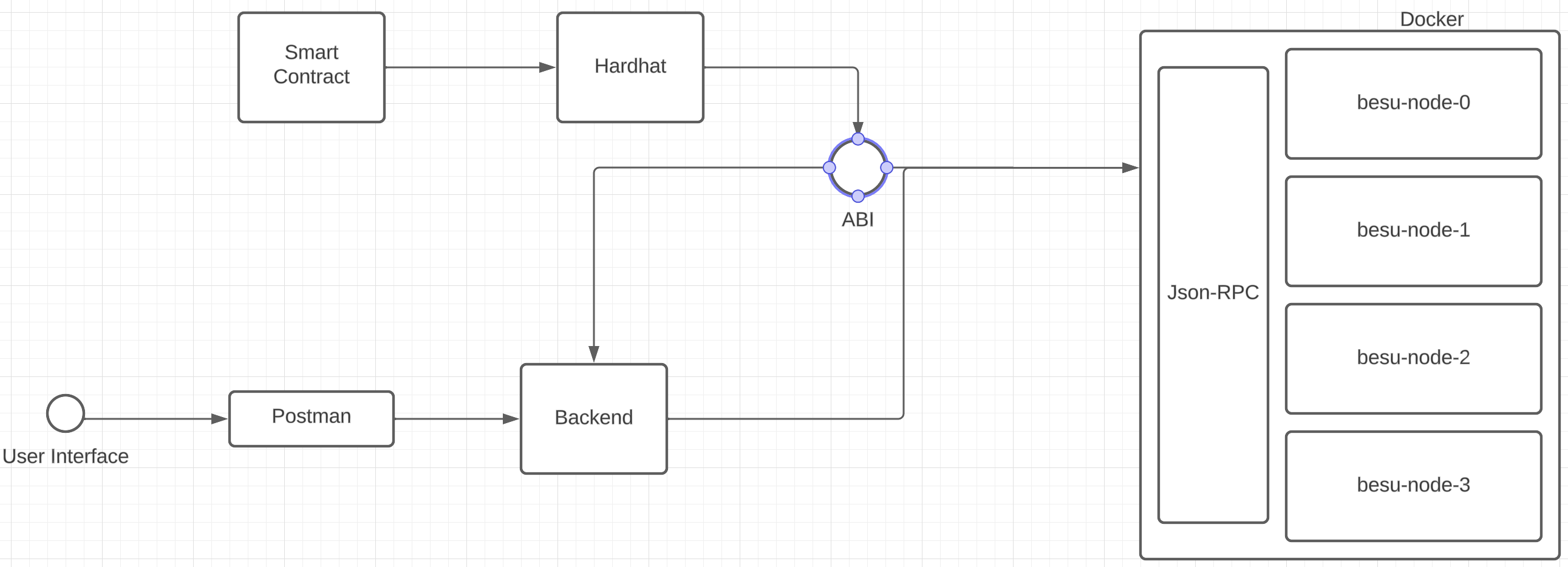}
  \caption{Illustration for the candidate architectural design with Hyperledger Besu.}
  \label{fig:diagrama}
\end{figure*}

\section{Proof of Concept}

This section shows the results of the Step 2, synthesis step, of Hofmeister's process. The output of this activity is a set of architectural alternatives to be assessed against the prioritized problems. Thus, it moves from the problem to the solution space \citep{HOFMEISTER2007}. 
In the next subsections, we delve into the details of the two raised candidate architectures conceived at this step.
\\ 
\vspace{-0.4cm}
\subsection{Candidate Architectural Solution 1: Hyperledger Besu}

The main objectives of the development of this candidate software architecture were to (i) Create a component-structured architecture, highlighting the use of smart contracts connected to a \textit{blockchain Besu Ethereum} network, (ii) Develop an API capable of verifying and validating news, enabling fact-checkers to distinguish between true information and \textit{fake news}, (iii) Demonstrate the practical feasibility of the project, using \textit{Hyperledger Besu} technology to implement the \textit{blockchain} network and (iv) Contribute to the creation of a more trustworthy and safe environment on social networks, promoting responsibility in the dissemination of information and combating the negative impacts of \textit{fake news}.

\noindent\textbf{Rationale for Using Hyperledger Besu.} \textit{Hyperledger Besu} is an open source Ethereum client developed by \textit{Pegasys}, a subsidiary of \textit{ConsenSys}. Besu stands out as a solution adapted for enterprise environments, covering use cases in both public networks as private with permissions. Its fundamental purpose is to provide businesses with an affordable and efficient platform for creating and managing decentralized applications (\textit{dApps}). Among its notable features are (i) support for networks using the Ethereum protocol, both public and private, (ii) corporate governance tools such as smart contract management, as well as features dedicated to scalability, and (iii) privacy protection and fluid integration with other networks blockchain, such as \textit{Quorum} \citep{Dalla2021}.

\textit{Hyperledger Besu} also relies on EVM (\textit{Ethereum Virtual Machine}) implementation, also providing flexibility. 
In terms of consensus, it offers a variety of algorithms, such as \textit{Proof of Stake}, \textit{Proof of Work} and \textit{Proof of Authority} (\textit{IBFT 2.0, QBFT and Clique}), providing options adaptable to the specific needs of each scenario. \textit{Hardhat}, presented as a toolkit, is a suite of tools designed to optimize the creation of smart contracts, providing a significant increase in productivity during development. Developed in \textit{Node.js}, it is essential to ensure that \textit{Node} is previously installed for its correct functioning \citep{Jain2023}.
The development environment encompasses several specialized components such as editing, compiling, debugging, and deploying smart contracts and decentralized applications (\textit{dApps}). The harmonious integration of these elements aims to create a complete and adequate development environment.
Among the various \textit{blockchain} platforms available, the choice of \textit{Hyperledger Besu} for this candidate architecture is justified by several distinctive characteristics \citep{hasan2020blockchain}, including (i) Modular Architecture, (ii) Security, (iii) Flexibility in Consensus Choice, (iv) Optimized Performance and Monitoring, (v) Scalability, and (vi) Support for Interoperable Protocols \citep{Alba2023, Fan2022}.
\\
\noindent\textbf{Architectural Design.} The solution has been meticulously developed to establish a local \textit{blockchain} environment, utilizing the \textit{framework} \textit{Hyperledger Besu} to compile and deploy smart contracts through \textit{Hardhat}. Additionally, a \textit{backend} was designed to interact with the \textit{blockchain}, and the features were validated and integrated with the \textit{Postman} tool. The vital process for executing the solution begins by configuring the local environment. The supplied \textit{bash} script configures the \textit{Besu} local network, employing the IBFT consensus algorithm. This environment consists of \textit{Docker} containers representing \textit{Besu} nodes, which are cleaned and configured before the process starts. Each node has its data directory, ensuring independence and local persistence.

\noindent\textbf{Configuration of Nodes in the Besu Network.} The \textit{blockchain} local environment is made up of several nodes, each one playing a specific role in the \textit{Hyperledger Besu} network. ``besu-node-0'' acts as the bootnode, coordinating the initialization and applying the IBFT consensus algorithm for reliability. ``besu-node-1, besu-node-2, besu-node-3'' are participating nodes, maintaining local copies of \textit{blockchain} and contributing to consensus, also using the IBFT algorithm. Communication occurs via the \textit{Docker} \textit{``besu\_network''} network, facilitating the exchange of information. Each node has its data directory, ensuring independence and local persistence, facilitated by \textit{Docker} volumes. Nodes expose RPC and HTTP interfaces for external interactions, such as smart contract compilation and deployment, using custom ports for efficiency. These features provide a robust environment for developing, testing, and effectively interacting with smart contracts on \textit{blockchain} \textit{Hyperledger Besu}, emphasizing stability and reliability.
\\\\
\noindent\textbf{Compilation and Implementation of Smart Contracts with Hardhat.} After the initialization of the local environment, \textit{Hardhat} takes on the responsibility of compiling and deploying \textit{Ethereum} smart contracts. The deployment script, exemplified by a contract called ``FakeNewsValidator'', is essential to this process. \textit{Hardhat} plays a crucial role in ensuring smooth integration between the local environment and the development cycle. Additionally, during compilation, \textit{Hardhat} generates essential folders such as \texttt{artifacts} and \texttt{cache} that store vital information including \textit{ABI (Application Binary Interface)} and  \textit{bytecode}.




\noindent\textbf{Development of the Backend and API for Querying Smart Contracts with Express.js.} The \textit{backend} of the solution was developed using the \textit{Express.js framework}, acting as an HTTP server for interaction with the \textit{Besu blockchain}. The \textit{Web3.js} library is incorporated into the code to enable efficient communication with the \textit{Besu} local network. \textit{Express.js} provides a foundation for building the \textit{backend}, allowing  to create a robust HTTP server. This server is essential to facilitate interaction between users and the \textit{blockchain}, enabling exposing data and functionalities through an \textit{API}. The \textit{API} developed in \textit{backend} offers an interface for querying deployed smart contracts. The /check-news/:newsId endpoint allows users to obtain specific information about a news story by validating input parameters to ensure the integrity of requests.
Furthermore, interaction with smart contracts is facilitated by the use of \textit{ABI (Application Binary Interface)} and the contract address. The \textit{ABI} is an interface that defines how a contract's methods can be called, while the contract address represents the specific location of that contract in the \textit{blockchain}. These elements are crucial for the correct communication and execution of transactions on \textit{blockchain}. \textit{Web3.js} uses this information to create calls to contracts and obtain the desired results, thus providing a simple and effective interface for end users.


\noindent\textbf{API Testing with Postman.} Validation of the solution's operation and integration between components are tested with the \textit{Postman} tool. This tool allows sending HTTP requests to \textit{endpoint /check-news/:newsId}, simulating user interaction with the \textit{API} developed. These tests contribute to verifying the integrity of \textit{API} responses and confirming the correct execution of smart contracts on the local \textit{blockchain}. 


 \begin{figure}[!hbt]
  \centering
  \includegraphics[scale=0.42]{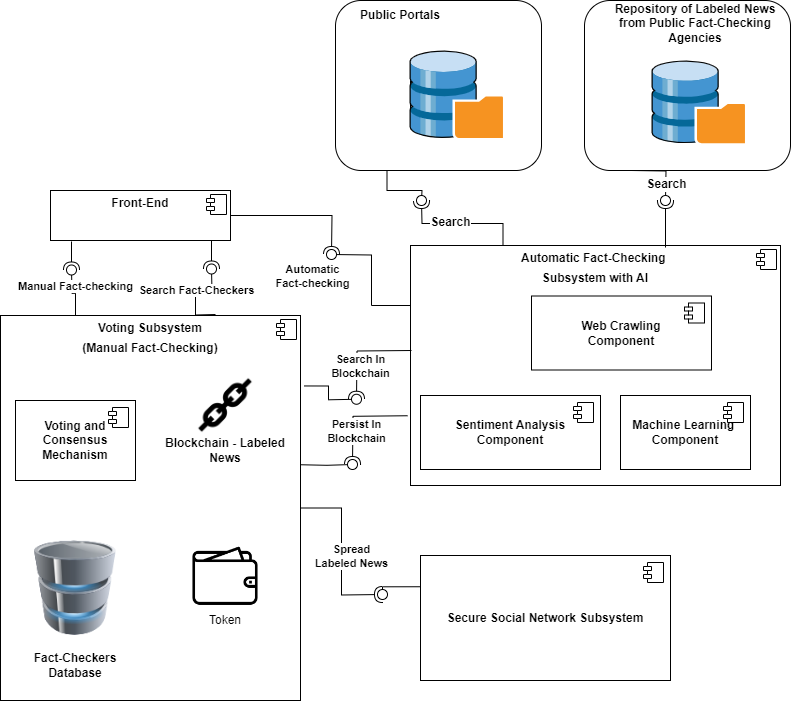}
  \caption{Diagram for the candidate architectural design with Hyperledger Fabric. Extracted from Private Technical Report \citep{techReport}.}
  \label{fig:diagrama2}
\end{figure}

\vspace{-0.5cm}
\subsection{Candidate Architectural Solution 2: Hyperledger Fabric}

This candidate architecture uses Hyperledger Fabric, instead.

\noindent\textbf{Rationale for Using Hyperledger Fabric.} Hyperledger Fabric is an open-source framework for private blockchains managed by the Hyperledger Foundation. The platform provides a fully-featured and modular architecture, allowing flexibility and expansion depending on the use case. One distinctive characteristic of Fabric is its concept of organizations, which makes it more suitable for enterprise situations. Organizations own components that interact in the network, peers, and orderers (as shown in Figure \ref{fig:diagrama2}). Each component has its responsibility within the lifecycle of a transaction. \citep{fabric-article}

Hyperledger Fabric supports smart contracts, often called chaincodes, written in general purpose languages, namely Go, Java and JavaScript \citep{fabric-article}. This lowers the development barrier and allows language-specific resources to be used in implementation, enabling more complex applications to be developed.
 Many blockchains implement a *Order-Execute* sequence for transactions, whereas Fabric implements a *Execute-Order-Validate* sequence, which enables much more secure and consistent blocks with finality. Hyperledger Fabric implements the concept of channels, where each channel has its own configuration, member organizations and hosts its ledger. Organizations can be part of multiple channels simultaneously, allowing a peer to host multiple ledgers without mixing different data scopes. This means that, unlike most blockchains, a node can be a part of multiple ledger-sharing channels, creating an extensive network \citep{fabric-article}. Fabric also implements what are called private data collections (PDCs). PDCs are environments for private data registration in a given chaincode. Only organizations that are members of the collection are given read access to data. The data is shared peer-to-peer via a gossip protocol and is never registered in the ledger. The ledger registers the hash of the private information to allow for proof of registry by parties involved \citep{fabric-article}. 

CC-Tools is a Hyperledger Labs open-source project and part of the toolkit for application development. CC-Tools provides features regarding asset, data type, event and transaction development. CC-Tools allows for more complex chaincodes while decreasing development time. This project supports chaincodes written in Go language and fully compatible with all major Fabric long-term support (LTS) versions. Hence, several features of Hyperledger Fabric can be listed as justifications for its choice, such as (i) Modular architecture, (ii) Smart contract flexibility, (iii) Security, (iv) Consensus, (v) Privacy, (vi) Performance \citep{Hyperledger2023Benchmarking}, (vii) Scalability, and (viii) Monitoring.

\noindent\textbf{Architectural Design.} Hyperledger Fabric provides all the features needed to create an entire network architecture for the proposed application. Chaincode implementation was done using the CC-Tools Hyperledger Labs project, where a CC-Tools-Demo repository was used as a base template for the implementation. This repository provides a working test network with three application organizations with one peer each and an ordering organization with one orderer, as shown in Figure \ref{fig:diagrama2}. The organizations are members of a channel where the chaincode proposed is instantiated. Three client APIs are also provided, tied to identities for each one of the application organizations.

\begin{figure}[hbt]
  \centering
  \includegraphics[scale=0.30]{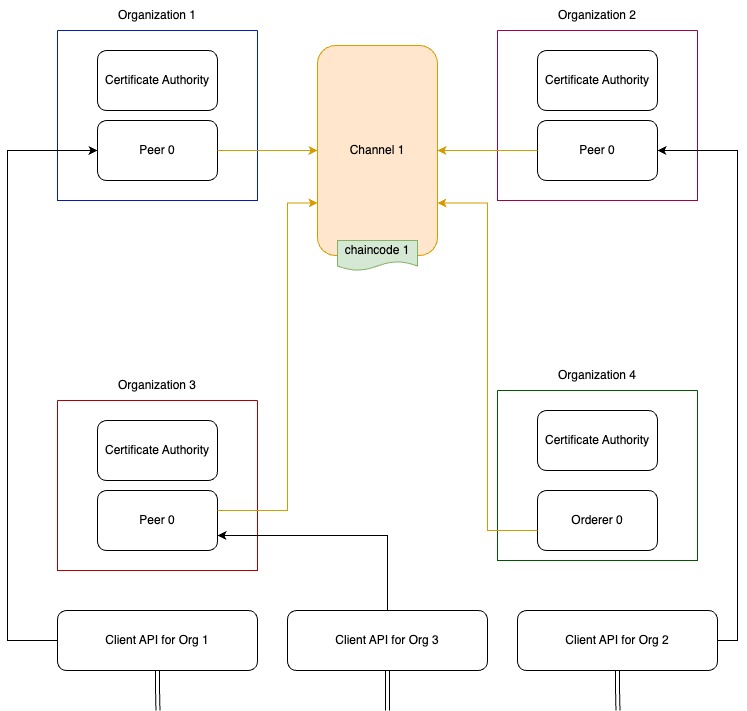}
  \caption{Diagram for the candidate architectural design with Hyperledger Fabric. Extracted from Private Technical Report \citep{techReport}.}
  \label{fig:diagrama}
\end{figure}

\noindent\textbf{Chaincode implementation.} The chaincode is implemented by defining the core assets and transactions within the structure provided by CC-Tools. Assets act as the data mapping that will be registered, that is, the actual properties that will be registered. Transactions implement the business logic that will act on ledger data, either reading from it or writing to it. 

\noindent\textbf{API implementation.} The client API is implemented in Go using the Fabric SDK package for interaction with the ledger. This API utilizes a certificate and private key issued by the CA of a trusted organization in the channel. These credentials authorize the API to interact with network. The API utilizes the Gin package for creating a REST API that exposes SDK functionality. For compatibility with older and newer versions of Fabric, the API implements endpoints that utilize a legacy SDK for transactions and a newer Gateway SDK that uses the peer Gateway Service, that facilitates transaction submission. 

 Inspired by Figure \ref{fig:scenario}, we depict Figure \ref{fig:diagrama2},  showing the component-and-connector (C\&C) view to represent our second candidate architectural solution, discussed as follows. 

\noindent\textbf{Automatic FactChecking Subsystem} - This is the subsystem responsible for conducting the web crawling in public portals, to search in the blockchain and conduct automatic analysis based on AI in the selected content. \textit{Matches ASR7}.

\noindent\textbf{Human Fact-Checkers Subsystem} - This subsystem involves the management of the database of human fact-checkers, the voting and consensus mechanism (materialized in a smart contract), and the blockchain itself with its interface to interoperate, externalize access and receive persistence demands. \textit{Matches SMR3, ASR4, ASR5, ASR6, ASR8, ASR9 and SMR7}.

\noindent\textbf{Voting and Consensus Mechanism} - This component manages the voting and consensus between the human fact-checkers to decide the legitimacy of the content of the news being analyzed. It is a component within the Fact-Checkers Subsystem. \textit{Matches ASR3}.

\noindent\textbf{Secure Social Network} - This component will be implemented to support a secure social network. \textit{Matches SMR6}.

\noindent\textbf{Services interface} - This part, not explicitly captured in the model of Figure  \ref{fig:diagrama2}, regards to the interfaces externalized for client consumers and software developers that can access and build applications over our public infrastructure, besides the hooks used to consult the external databases of public portals and social networks. \textit{Matches ASR1 and ASR2}.



\begin{figure*}[hbt]
  \centering
  \includegraphics[scale=0.50]{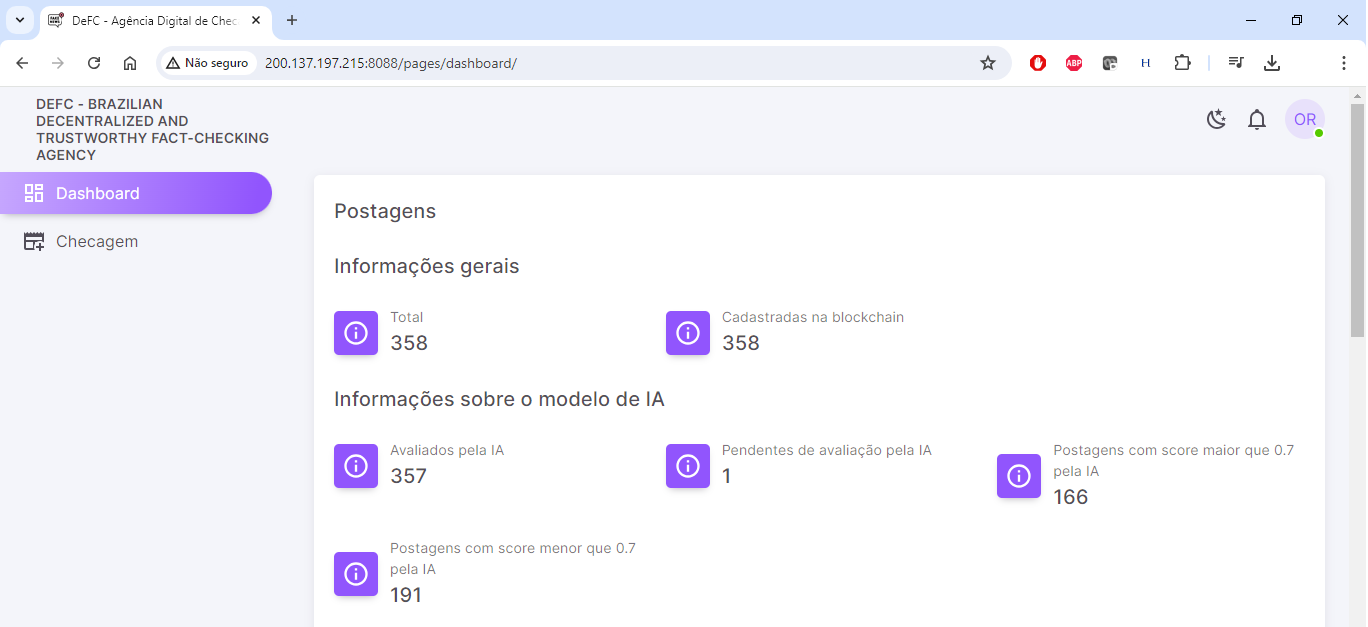}
  \caption{A screenshot of the interface of the tool conceived on the selected architectural design.}
  \label{fig:screenshot}
\end{figure*}

\noindent\textbf{Decision.} Given the two candidate architectures and based on the prioritized requirements, the second solution was chosen, based on Hyperledger Fabric. Fabric's modular architecture allows for a high degree of customization. This modularity enables organizations to tailor the blockchain to their requirements, including endorsement policies, membership services, and network setup. Such flexibility is paramount for applications with unique needs or operating in highly regulated industries.
Furthermore, Hyperledger Fabric provides enhanced transaction confidentiality through its architecture, which supports the execution of transactions within a private context. This is facilitated by its unique approach to channels, where a subset of participants can conduct transactions privately, a particularly appealing feature for applications requiring confidentiality in their operations. Besu offers privacy features, such as private transactions and privacy groups. However, these are built on top of a platform initially designed for public network compatibility, making Fabric's privacy features more robust for some applications.
Additionally, Hyperledger Fabric supports using general-purpose programming languages for crafting smart contracts. Unlike Besu, where the EVM specifications constrain smart contract capabilities, Fabric's chaincode can exploit the comprehensive functionalities offered by languages like Go and their extensive libraries. This distinction allows for a broader and more versatile development environment in Fabric, catering to complex enterprise needs beyond the scope of what Solidity and the EVM can provide on Besu. 
Once the candidate architectural solution was selected (and although the another solution was also implemented as a prototype), the proof of concept was then conceived, as follows.

\begin{figure}[hbt]
  \centering
  \includegraphics[scale=0.34]{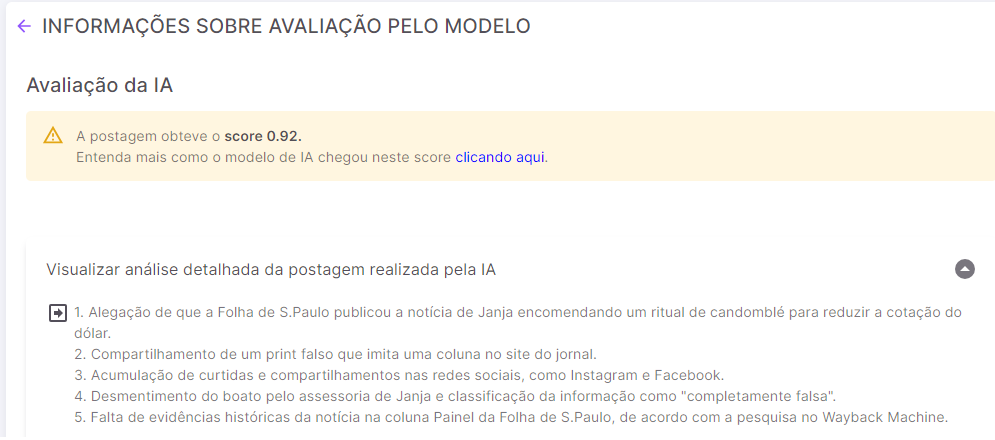}
  \caption{A screenshot of the AI classification for a single news with explainability.}
  \label{fig:screenshot2}
\end{figure}

\subsection{Prototype}
The prototype was named Brazilian Decentralized And Trustworthy Fact-checking Agency (DEFC). The screenshot shows, in Portuguese, one of the main menu items we have (a dashboard). The dashboard displays general strategic information for the public entities that are interested in those results, such as the total of news registered and the total registered in Blockchain, the number of news evaluated by the AI mechanism, those that still demand evaluation, the number of posts under analysis that has a score greater than 0.7 and were evaluated by the AI. The closer the score is to 1, the more likely the news is false, as priorly stated. The another main functionality is the fact-checking, which supports the fact-checker to gather evidence to judge the veracity of one or more posts under analysis, following the pipeline illustrated in Figure 1. Once a established minimum number of fact-checkers decide on the status of a post, this is revealed for all of them, stored in the blockchain, and made available for the competent authorities.

The user can assume two different roles in this tool: (i) a fact-checker, i.e., the human user who will assess the fakeness of the posts under analysis; and (ii) a curator, who will be the gatekeeper that can authorize new fact-checkers to join the board. The latter can also analyze the entire platform in a panoramic perspective.

Step 3 of Hofmeister's Process, Evaluation, is discussed in the next section.

\section{Evaluation and Discussion}
\label{sec:discussion}


An evaluation of the architecture was performed inspired in the study of Abreu et al. (2020) \citep{abreu2020}, based on User Validation Scenario, Technical Validation and Viability of the Blockchain Use, as follows. The steps are summarized due to space restriction.
\\\\
\noindent\textbf{6.1. User Validation Scenario.} A typical user of this tool is the fact-checker, i.e., the professional dedicated to the analysis of suspect posts in social media. As such, the credentials of the tools were made available for two fact-checkers of two different agencies: Aos Fatos\footnote{\url{https://www.aosfatos.org/}} e 
Boatos.org\footnote{\url{https://www.boatos.org/}}. The fact-checkers could register novel news to be analyzed, obtain the score assigned by the AI mechanism and gather evidence and report on their beliefs abou the fakeness of the news just registered in the system. The score delivered by the AI subsystem is accompanied by the explainability, i.e., the reasons that led it to conclude about that post, as shown in Figure \ref{fig:screenshot2}.

After the use, an interview was conducted with them to assess the usage of the tool when compared to the current process. The participants cleared stated that gains can be obtained from the use of that solution, including (i) \textbf{time/productivity}, since the AI mechanism already gather some evidences and scores the news; the fact-checker only needs to complement the evidences or even agreeing with the results, accelerating the process. One participant highlighted that this is essential, since the time a fake news is being spread without being classified as such can be determinant for the impacts of it; and (ii) \textbf{reliability}, since not only a single fact-checker will work on each suspicious news, but a pool of them, supported by a consensus mechanism and the blockchain infrastructure.
\\\\
\noindent\textbf{6.2.Technical Validation.} As shown in Figure \ref{fig:screenshot}, more than 300 posts were analyzed by the AI, labeled and stored in the blockchain. Abreu et al. (2020) \citep{abreu2020} mentions the technical challenges posed by the use of blockchain in a commercial solution, as listed by Ciccio et al. (2020) \citep{di2020business}. The technical challenges include  
smart contracts (reflecting monitoring capabilities), oracles (identifying data sources) and data monitoring (balancing between data inside and outside the chain). Several other related challenges exist for each of these categories. 

Abreu et al. (2020) \citep{abreu2020}, citing Ciccio et al. (2020) \citep{di2020business}, mentions that the challenges related to smart contracts are (i) Monitoring Transparency: anybody can see the logic and data behind monitoring,
data required for smart contracts must be provided and confidential data should not be shared or be protected; (ii) Observability: to access monitoring data, smart contracts must implement mechanisms to expose them, for instance, private variables can not be easily accessed; and (iii) Lack of reactivity: smart contracts can not
directly invoke external services. For challenges on oracles: (i) Time management: the blockchain lacks the notion of time, and timers can not expire by themselves, an
external triggering event is always required; (ii) Reliability: oracle breaks the decentralizsed trust and to compensate, several oracles should be used; and (iii) Flexibility: oracles are bound to smart contracts because a compromised oracle can not be replaced during execution, nor new capabilities can be introduced. Finally, as challenges for data management: (i) Data quality: the
quality of the data influences the monitoring results, for example, poor quality data sources compromise the monitoring and once stored, incorrect data cannot be changed; (ii) Data size: the cost of the blockchain is proportional to the amount of data stored, for
example, in public blockchains, monetary cost (cryptocurrency), in private blockchains, overhead on the platform, and also storing data off-chain; and (iii) Side effects: most blockchains are prone to forks, and this may cause contradictory information or interoperability issues.

Regarding those issues, about transparency, only the fact-checkers identities are confidential data that should not be shared or be protected; about observability, there is an additional layer that restricts the access only to the data that should be accessed by others, i.e., the labeled news and their history during the process; and analogously to the prior justification, the smart contracts do not
directly invoke external services (Lack of reactivity); an additional layer was conceived for that purpose. About oracles, we did not have resources in the blockchain infrastructure to measure the exact time of a transaction. The approximate time of transactions was measured without the need for a policy trust between nodes. Finally, as about data management, (i) about data quality, the news were only labeled after the consensus run by the smart contracts and with results from the fact-checkers; then, the prototype has quality and correct data; (ii) about data size, only textual information was stored in the blockchain; the other information was replicated in ElasticSearch; and (iii) side effects are avoided, since the infrastructure was outsourced, and they implement mechanisms to avoid that.
\\\\
\noindent\textbf{6.3. Viability of the Blockchain Use.} For a viability analysis, the ten steps
proposed in \citep{pedersen2019ten} were used the Ten-step Path for Blockchain Validation, which support to decide whether or not to use blockchain technology in our prototype, as perfomed by \citep{abreu2020}. The ten steps consist of questions that must be made before deciding to use blockchain. As a general rule, Pedersen et al. (2020) \citep{pedersen2019ten} recommends a blockchain is feasible to use if five questions of the applied questionnaire are answered as “Yes”. We discuss each one, as follows.

Currently, the process of fact-checking is often questioned because it is performed by single fact-checkers in their agencies.  Fact-checkers can have bias in their judgement and the results should be stored in a single shared space. Multiple fact-checkers from a diversity of opinions and agencies should be involved to check each news. Then, this brings \underline{Yes} for the questions: \textbf{Need for a shared common database?; Multiple parties involved? Involved parties have conflicting interests/trust issues?} 
If the tool is owned by the government and the fact-checkers are also from the same entity, this can be target of doubts and attacks. Then, this is a \underline{Yes} for \textbf{Parties can/want to avoid a trusted third parties?} 
Finally, all the history of processing and analyzing each news should be stored indefinetively, allowing traceability and auditing, which poses a \underline{Yes} for \textbf{Need for an objective immutable log?}\footnote{The other questions are more technical than business-oriented. Since we already have the Yes for 5 of them here, the others will not be shown due to space restrictions.} 
\\\\
\noindent\textbf{6.4. Brief Discussion and Threats to Validity.} Other blockchain technologies could have been considered as candidates for our Proof of Concept (PoC). However, both Hyperledger Besu and Hyperledger Fabric presented desirable characteristics that weighted the decision towards them. Hyperledger Fabric offers a modular and configurable architecture that enables a high degree of privacy, scalability, and flexibility in transaction management, essential for meeting enterprise-level applications' diverse needs. Its support for smart contracts, known as chaincode, allows for developing complex business logic that can be securely executed within the blockchain network. 


While blockchain offers potential benefits for creating a transparent and immutable record of news articles, its application in fighting fake news can face important challenges, particularly regarding feasibility, scalability, and cost. About feasibility, the success of a blockchain-based system depends some important technical concerns, including its large-scale adoption, the hardware infrastructure to deploy it with elastic capabilities to respond to the increase of resource demands and even a pool of fact-checkers that could contribute to the voting process. All these concerns lead to other two important issues: scalability and costs. About scalability, such a system should be prepared to support a large degree of transaction throughput, data storage, and network load. Since the system is still a proof-of-concept, this is not a concern. Once it is made available to the population, a load balance is need both to receive the queries and to process them, maybe in a queue. Services redundance is also needed, which is supported due to the technology stack used even in the PoC version. Moreover, by the blockchain nature, more nodes can be created to accordingly support an increasing demand in the services, splitting the demands in groups of nodes inside the blockchain. Finally, about costs, we could consider three types: (i) infrastructure, (ii) transactions and (iii) fact-checkers payment. In all the cases, this would not be a considerable problem, since this is a government initiative. Then, the government could (i) acquire the necessary hardware to deploy the blockchain network accordingly, (ii) pay for the gas eventually needed to emit the tokens and (iii) pay the fact-checkers, converting tokens into currency, as in other already existing initiatives, such as the bases for educational institution assessment.

\noindent\textbf{Threats to Validity.} This research could have been affected by different factors \citep{wohlin2012experimentation}, and we discuss them, as follows.
\\
\textbf{Internal Validity.} Internal validity concerns to the validity within the given environment and the reliability of the results. As stated by Abreu et al. (2020), The network environment can fluctuate in terms of latency, execution time for queries and transactions, block validation, disk space, and other factors. This variability can be problematic if the application requires immediate results. To address this, extensive testing of the application can be conducted to evaluate quality attributes and ensure compatibility with the environment. Another issue is that even if data is correctly registered on the blockchain, it can still be incorrect. Since the blockchain is immutable, this erroneous data remains recorded. To rectify this, a new correction record must be added, and the application must be designed to manage such cases.
\\
\noindent\textbf{External Validity.} External validity concerns the extent to which the results of a case study can be generalized. This study involved only two fact-checkers, raising questions about scalability. Additional research is needed to test the prototype in broader contexts, involving more fact-checkers and processing a larger number of posts for registration and labeling.
\\
\textbf{Conclusion Validity.} Conclusion validity pertains to the relationship between the treatment and the outcome. The evaluation conducted involved a small number of participants and relied on subjective questions. Additional research involving a larger group of fact-checkers is necessary for more robust results.
\\
\noindent\textbf{Construct Validity.} Construct validity focuses on the link between theory and observation. In this work, we only considered opinions from individuals at two fact-checking agencies. These individuals lacked technical expertise in blockchain, limiting the depth of their responses to their knowledge of the fact-checking process. The data used were sourced from actual news portals. Moreover, the application was not deployed for production use but was only utilized for this research. There is a need to evaluate the application with real, large-scale data, considering impacts on quality attributes and involving multiple users, to truly demonstrate its effectiveness and determine whether investing in blockchain is worthwhile.

\section{Final Remarks and Future Work} 
\label{sec:FR}

The main contribution of this paper was to report the creation of a blockchain-based software architecture of a solution to fight fake news dissemination, within the project scope of a dApp PoC \citep{techReport}. The architecture was systematically conceived following the canonical architectural framework of Hofmeister \citep{hofmeister2007general}. We assessed two different candidate architectures based on different blockchain technologies and chose one to be evaluated using simulation. This study brings insights and lets the concerns be accordingly recorded so that iterations over the architecture can enhance the satisfaction of other requirements. This work demonstrates a practical approach to creating local blockchain development environments. Possible extensions include incorporating authentication and authorization mechanisms, deeper analysis of communication between nodes, and exploring advanced features of Hyperledger Besu.

Nevertheless the focus of the architecture proposed herein was not on the availability of the services for the public or other entities, we already have an evolved proposal of it to provide an external endpoint exposed as an API for news agencies, social media platforms and even a service for the population to use and collect information regarding the validity of the information they want to check \citep{brasnamValdemar2024}. Agencies and social media platforms could use the API to award badges to verified information, giving their users more transparency about the news they consume whilst the population could use the service to check suspicious content they find in their daily lives. We expect the advances achieved here can be reproduced/replicated in several countries to combat fake news dissemination or even in other domains, such as Smart cities \citep{NetoKassab2023}.

\section*{Acknowledgements}

 We thank ANATEL (the Brazilian National Telecommunications Agency), who supported this research.

\bibliographystyle{ACM-Reference-Format}
\bibliography{sample-base}


\end{document}